\documentclass[letter]{IEEEtran}
\usepackage{amsmath,amssymb,amsfonts}
\usepackage{graphicx}
\usepackage{cite}
\usepackage{balance}
\usepackage[keeplastbox]{flushend} 
\usepackage{epstopdf} %converting to PDF
\usepackage{algorithm}
\usepackage[noend]{algpseudocode}
\usepackage{tikz}
\usetikzlibrary{automata, positioning}
\newcommand{\RN}[1]{%
\textup{\uppercase\expandafter{\romannumeral#1}}
}
\begin{document}

\title{A Weather-Dependent Hybrid RF/FSO Satellite Communication for Improved Power Efficiency}
\IEEEoverridecommandlockouts 
% \author{Authors}
\author{Olfa Ben Yahia, Eylem Erdogan, \textit{Senior Member}, \textit{IEEE}, Gunes~Karabulut~Kurt, \textit{Senior Member}, \textit{IEEE}, \\ 
Ibrahim Altunbas, \textit{Senior Member}, \textit{IEEE}, Halim Yanikomeroglu, \textit{Fellow}, \textit{IEEE}%
\thanks{O. Ben Yahia, G. ~Karabulut~Kurt and I. Altunbas are with the Department of Electronics and Communication Engineering, Istanbul Technical University, Istanbul, Turkey, (e-mails: \{yahiao17, gkurt, ibraltunbas\}@itu.edu.tr).}%
\thanks{E. Erdogan is with the Department of Electrical and Electronics Engineering, Istanbul Medeniyet University,  Istanbul, Turkey, (e-mail: eylem.erdogan@medeniyet.edu.tr). }%
\thanks{G.~Karabulut~Kurt is also with the Department of Electrical Engineering, Polytechnique Montréal, Montréal, QC, Canada (e-mail: gunes.kurt@polymtl.ca). }% 
\thanks{H. Yanikomeroglu is with the Department of Systems and Computer Engineering, Carleton University, Ottawa, ON, Canada, (e-mail: halim@sce.carleton.ca).}}
\maketitle
\begin{abstract}
Recent studies have shown that satellite communication (SatCom) will have a fundamental role in the next generation non-terrestrial networks (NTN). In SatCom, radio-frequency (RF) or free-space optical (FSO) communications can be used depending on the communication environment. Motivated by the complementary nature of RF and FSO communication, we propose a hybrid RF/FSO transmission strategy for SatCom, where the satellite selects RF or FSO links depending on the weather conditions obtained from the context-aware sensor. To quantify the performance of the proposed network, we derive the outage probability expressions by considering different weather conditions. Moreover,
asymptotic analysis is conducted to obtain the diversity order. Furthermore, we investigate the impact of non-zero boresight pointing errors and illustrate the benefits of the aperture averaging to mitigate the effect of misalignment and atmospheric turbulence. Finally, we suggest effective design guidelines that can be useful for system designers. The results have shown that the proposed strategy performs better than the dual-mode conventional hybrid RF/FSO communication in terms of outage probability offering some power gain.
\end{abstract}
%*********************************%
\section{Introduction}
%*********************************%
Next-generation communication networks will integrate space communication with terrestrial infrastructure for heterogeneous networks. 
%Numerous projects including SpaceX’s Starlink and Amazon’s Project Kuiper consider providing 5G/6G connectivity around the world via satellite mega-constellations \cite{9380673}. 
 {In the near future, low Earth orbit (LEO) satellites, which are able to cross over the continents within few minutes, are expected to play a key role in the next generation networks, due to their intrinsic features. LEO satellites introduce agility to the network and enable high throughput broadband services with low latency \cite{9210567}.
To address the traffic demand of these services, }the use of free-space optical (FSO) communication has attracted considerable attention from both the research community and industry for satellite communication (SatCom) networks. In fact, the FSO technology provides several advantages such as high data rates, high bandwidth, and low cost \cite{9296829}.
\begin{figure}[!t]
  \centering
    \includegraphics[width=2.8in]{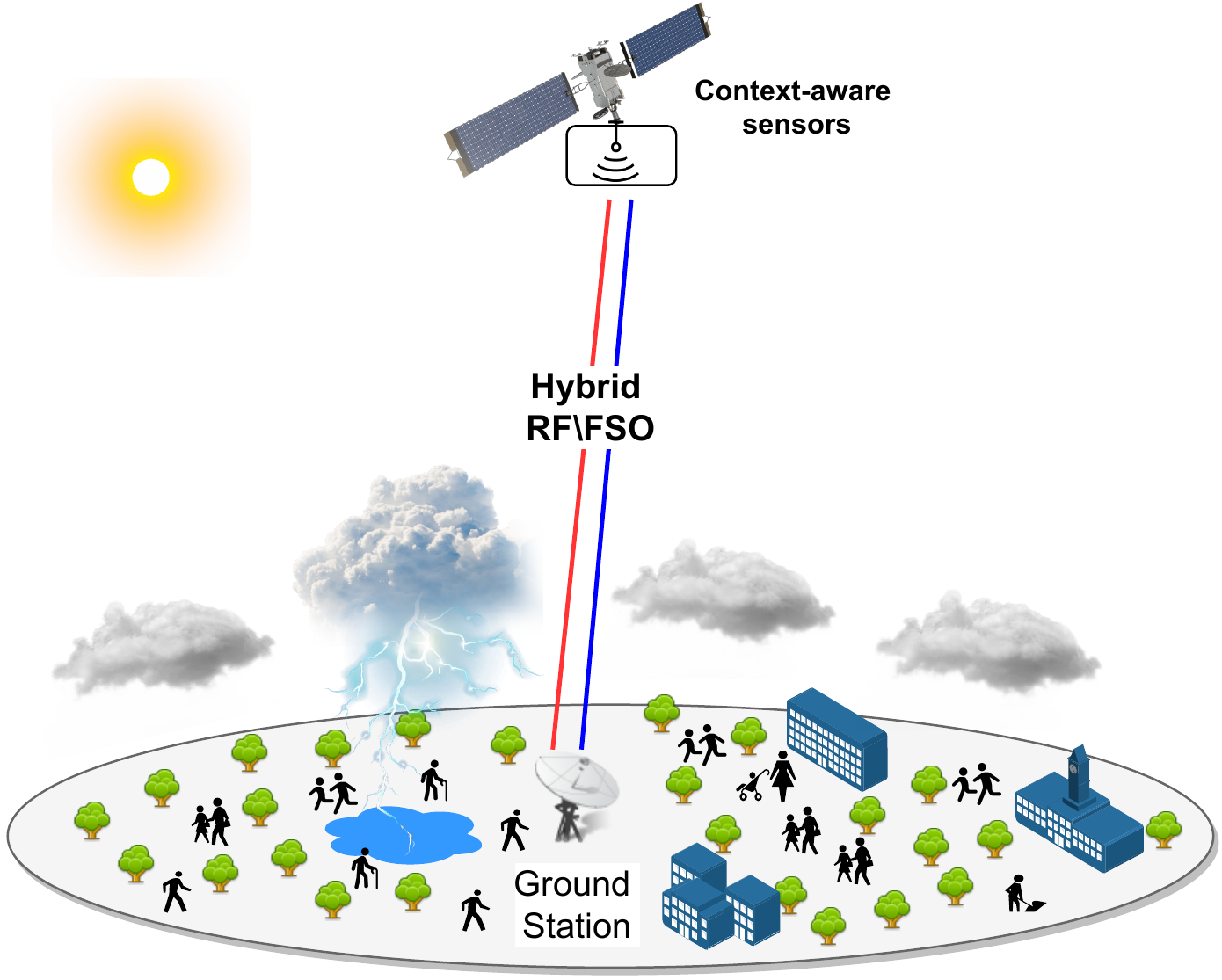}
    \caption{Illustration of weather-dependent hybrid RF/FSO SatCom. }
  \label{fig:model}
\end{figure}  
Furthermore, FSO communication requires perfect line-of-sight (LOS) connectivity due to its narrow transmit bandwidth.
Despite its great potential, FSO communication can experience reliability and availability issues due to atmospheric turbulence-induced fading, and sensitivity due to weather conditions including dense clouds, fog, and snow \cite{viswanath2015}. However, the impact of rain on FSO transmission is less significant compared to fog. Another major impairment for FSO links is the pointing error, which implies the possible displacement of the beams due to large distances, wind, pressure, and temperature variations \cite{2018outage}. Thereby, several techniques have been proposed in the literature to mitigate these adverse effects \cite{touati}. The use of radio frequency (RF) and FSO technologies together is an attractive solution to address this gap \cite{soleimani2015generalized}. This combination consists of two possible configurations. The first one is the so-called mixed RF/FSO, in which RF communication is used at one hop and FSO communication at the other in a dual-hop or relay-assisted networks \cite{soleimani2015generalized},\cite{ashrafzadeh2020unified}.
The second combination is hybrid RF/FSO communication, in which RF link is incorporated in parallel with FSO link. In fact, the FSO and RF links are sensitive to non-overlapping conditions and considered to have a complementary nature \cite{weather}. So far, different approaches have been proposed in this context by considering soft-switching or hard switching techniques \cite{kazemi}. In soft switching, both links might be active depending on their availability, whereas in hard switching, only one link can be active at a certain time. In the current literature, only a few papers have investigated hybrid RF/FSO communication for satellite systems [10,12-14] and, neither of them investigated the practical effects of weather conditions. Furthermore, pointing error, which is a critical metric in FSO communication, is barely investigated in hybrid RF/FSO configuration. \nocite{siddharth2020}\nocite{9446153}\nocite{swaminathan2021haps}\nocite{ben2021haps}

Different from the current literature, we provide a reliable communication strategy for the downlink SatCom by using hybrid RF/FSO communication considering soft-switching technique according to weather characteristics. More precisely, the proposed strategy enables us to switch between RF and FSO links depending on the weather conditions which can be obtained by using a context-aware sensor node on the satellite. In this concept, we assume that the satellite detectors are able to scan the weather conditions of the atmosphere within the LOS continuously. Then, based on the extracted information, the satellite determines to activate the most favorable channel immediately.
By doing so, the overall performance of the proposed scheme can be enhanced as the LEO satellite is aware of the weather conditions. For the proposed setup, FSO communication is modeled with exponentiated Weibull (EW) fading as it provides a better fit than all other models for different aperture sizes \cite{barrios2013}, whereas shadowed-Rician distribution is used to model the RF communication. Furthermore, we study the effect of atmospheric turbulence, scattering, and non-zero boresight pointing errors, and consider the impact of shadowing on communication performance. We also highlight the impact of aperture averaging to mitigate the effect of turbulence-induced fading and the misalignment of the transmitting beam to achieve better directivity. Outage probability (OP) expressions are derived for the proposed configuration. In addition, an asymptotic performance analysis is conducted and the achievable diversity gain is obtained.
Finally, we compare our proposed strategy with the conventional dual-mode RF/FSO soft-switching communication and provide interesting design guidelines.

The remainder of the paper is organized as follows. Signals and system models are presented in the following section. In Section $\RN{3}$, our proposed scheme is presented and the expressions of outage probability are derived. Numerical results are outlined and discussed followed by some design guidelines in Section $\RN{4}$, and Section $\RN{5}$ concludes the paper.
%*********************************%
\section{Signals and System Model}
%*********************************%
In this paper, we consider a downlink SatCom, where an LEO satellite aims to communicate with a ground station (GS) where both RF and FSO channels are used in parallel to transmit identical rate information, based on their availability as shown in Fig. 1. More precisely, the LEO satellite, which is deployed with context-aware sensors,  can detect the weather conditions on the ground continuously and adapt to the situation immediately. Thereafter, it uses the most favorable channel or both channels at the same time to transmit the information and accordingly adapts the transmission power. In the state where both channels are activated, the received signals are combined at the GS using the selection combining (SC) technique to maximize the overall performance.  Although the maximum ratio combining (MRC) technique provides better performance, we consider SC due to its practical relevance. Table $\RN{1}$ summarizes the proposed scheme.
The signal models for both channels are given in the following subsections.

\small
 \begin{table}[!t]
   \renewcommand{\arraystretch}{1.1}
   \centering
\caption{Proposed Scheme} 
\label{tab1}
\begin{tabular}{|p{0.5cm}|p{1.4cm} |p{2.6cm}|p{1.7cm}|  }
% \multicolumn{2}{|c|}{\textbf{FSO link}} \\
\hline  \textbf{State ($k$)} & \textbf{Description}  & \textbf{Observation} & \textbf{Transmission strategy}\\
 \hline \hline \textbf{0} & Thin Cloud & Both FSO and RF channels exhibit a higher performance & Dual (SC) \\
\hline \textbf{1} & Rain  & Only the FSO communication is favorable with full power  & FSO\\
\hline \textbf{2} & Fog  & Only the RF channel is available with full power & RF\\
\hline 
\end{tabular}
\end{table}
\normalsize
\vspace{-0.15cm}
\subsection{FSO Channel}
%************************%%
The received optical signal\footnote{For FSO transmission, we consider intensity modulated direct detection (IM/DD) as it is largely used in practical urban networks owing to its simplicity and low cost \cite{yi2015free}.} detected by the photodetector of the GS is modeled as $y =  \zeta \sqrt{P_1} I {x} + n$. $0\leq \zeta\leq1$ denotes the optical-to-electrical conversion coefficient, $P_1$ is the transmit power of the satellite through the FSO link, $I$ indicates the irradiance of the channel where $I=I^t I^p I^a $, where $I^t$ indicates the atmospheric turbulence-induced fading \cite{erdogan2020}, $I^p$ is the pointing error and, $I^a$ defines the atmospheric attenuation that includes Mie scattering and geometrical scattering \cite[eqn. (10)]{erdogan2020}. In clear weather conditions, the geometrical scattering is mainly caused by thin cirrus clouds and the attenuation coefficient is expressed as given in \cite[eqn. (8)]{erdogan2020}, whereas under rainy conditions, the extinction coefficient can be expressed as $\Theta=1.067\mathcal{R}^{0.67}$ \cite{kazemi}, where $\mathcal{R}$ is the rainfall rate in (mm/h).= ${x}$ is the transmitted signal and $n$ is the zero-mean additive white Gaussian noise with one-sided power spectral density $N_{0}$. Thus, the instantaneous signal-to-noise ratio (SNR) at GS can be expressed as
\begin{align}
   {\gamma}_{FSO} =\frac{ \zeta  P_1 I^2}{N_{0}} = \overline{\gamma}_{FSO}I^2,
\end{align}
where $\overline{\gamma}_{FSO}=\frac{ \zeta  P_1 }{N_{0}}$ is the average SNR of the FSO link and $\mathbb{E}[I^2]=1$. In FSO communication, we consider the combined effects of atmospheric turbulence, non-zero boresight error and jitter. Thereby, the cumulative distribution function (CDF) of $I$ can be given as \cite{2018performance}

\vspace{-0.2cm}
\small
\begin{align}
& F_I(I)=\frac{\alpha g^2}{\beta} \exp \Big(- \frac{s^2}{2\sigma_s^2}\Big) \sum_{i=0}^{\infty} 
\frac{(-1)^i \Gamma(\alpha)}{(1+i)!\Gamma(\alpha-i)}  \nonumber\\ 
&\times \sum_{j=0}^{\infty} \frac{(\frac{s^2 g^2}{2\sigma_s^2 \beta})^j}{j!}
G_{j+2,j+3}^{j+2,1} \Bigg( T_3(i) \Big| \begin{array}{c}1,
\overbrace{T_1+1 ,...,T_1+1}^{(j+1) \text{terms}} \\
1,\underbrace{T_1 ,...,T_1}_{(j+1) \text{terms}},0
\end{array} 
\Bigg),
\end{align}
\normalsize
where $\alpha$, $\beta$ denote the shape parameters \cite{barrios2013}, $g=w_{eq}/(2\sigma_s)$ is the ratio between the equivalent beam which is given as $w_{eq}^2=w_z^2 \sqrt{\pi} \text{erf}(v)/(2v e^{-v^2})$ and the jitter standard deviation $\sigma_s$, $\eta$ is the scale parameter \cite{barrios2013}, $s$ presents the boresight value, $T_3(i)=(1+i) \Big[  \frac{I}{(\eta A_0 )}\Big]^\beta$ with $A_0=[\text{erf}(v)]^2$ is the gathered optical power for zero difference between the optical spot center and the detector center, and $\text{erf}(\cdot)$ is the error function. Moreover, $v=\sqrt{\pi/2a}/w_z$ is the ratio of the aperture radius $a$ and the beam-width $w_z$ at the distance $z$ with $w_z=\theta z$, where $\theta$ denotes the transmit divergence angle, and $T_1=g^2/\beta$ \cite{yang2014free}. Finally, $\Gamma(\cdot)$ is the Gamma function and $G_{p,q}^{m,n} \Big( x\hspace{0.1cm} \Big| \begin{matrix} a_1,...,a_p\\  b_1,...,b_q \end{matrix} \Big)$ is the Meijer G-function \cite{Wolform}. In the absence of pointing error, the CDF of $I$ is given as \cite[eqn. (12)]{erdogan2020}. The parameters $\alpha$, $\beta$, and $\eta$ depend on the zenith angle of the satellite $\xi$, the wind speed level $v_g$ expressed in m/s, the wavelength $\lambda$, the refractive index $C_n^2$, and the aperture diameter $D_G$ \cite{erdogan2020}.
%************************%%
\vspace{-0.2cm}
\subsection{RF Channel}
%************************%%
Let $x$ denote the transmitted signal with power $P_{2}$ from the satellite to the GS through the RF link. The received signal at GS can be expressed as $ y =\sqrt{P_2 h_l} h  x + n $. $h$ denotes the channel response of the path from the satellite to the GS that follows the shadowed-Rician fading \cite{6605500}, $h_l$ is the path loss modeled as $h_l [dB]=G_T +G_R -20\log_{10} \left( \frac{4 \pi L}{f}\right)  -\omega_{oxy}L -\omega_{rain}L $ \cite{kazemi}, where $G_T$ and $G_R$ are the gains of the transmitting and receiving antennas respectively, $f$ is the RF wavelength, $L$ is the propagation distance in km, $\omega_{oxy}$, $\omega_{rain}$ are the attenuation coefficients due to the oxygen and rain scattering respectively \cite{touati}, and $n $ is the additive white Gaussian noise (AWGN) with one-sided power spectral density $N_0$.
Therefore, the instantaneous SNR at the GS can be written as
\begin{align}
\gamma_{RF}=\frac{P_{2} h_l |h|^2}{N_0} = \overline{\gamma}_{RF}|h |^2,
\end{align}
where $\overline{\gamma}_{RF}=\frac{P_{2} h_l}{N_0}$ indicates the average SNR of the RF link with $\mathbb{E}[|h|^2]=1$.

The main attenuation factor for the RF communication is the rain, where it rises linearly with the rainfall rate. Thus, the rain attenuation coefficient $\omega_{rain}$ (dB/km) can be expressed based on the rain rate $\mathcal{R}$ as $\omega_{rain}= k_r \mathcal{R}^\varrho$, where $k_r$ and $\varrho$ depend on the channel frequency $f$ \cite{ITUrain}.
The CDF of the squared envelope of shadowed-Rician channel $|h|^2$ can be expressed as \cite{paris2010closed}
\begin{align}
 F_{|h|^2} (h)= \mu h \Phi_2 \Big(1-m,m;2;-\nu h ,- \vartheta h\Big) ,
   \label{Eq4}
\end{align}
where $\mu =\frac{1}{2b }(\frac{2 b  m }{2 b  m  + \Omega })^{m }$, with $m $ representing the Nakagami-m severity parameter of the corresponding link and $\Omega $ and $ 2b$ are the average power of the LOS component and multi-path component, $\nu=\frac{1}{2b }$, and $\vartheta=\frac{m}{2bm+\Omega}$. Finally, $\Phi_2$ is the bivariate confluent hypergeometric function defined in \cite[Section 9.26]{2014table}.
%******************************% 
\vspace{-0.2cm}
\section{Performance Analysis}%%
%\vspace{-0.2cm}
%*********************************%
In this letter, we propose a new scheme in which a 
LEO satellite adapts its power and switches between RF and FSO links depending on the weather conditions obtained from the satellite sensors as shown in Table~$\RN{1}$. To this end, the overall system performance can be improved. 
Considering $P_t$ is the total transmit power of the satellite, we allocate equal transmit power on both RF and FSO links where $P_1=P_2=P_t/2$ when both links are used as given in State-0.  However, power allocation scheme can be adopted here to adjust the power levels if needed. For this state, we assume clear weather conditions (thin cloud), where both RF and FSO channels provide higher performance. For \mbox{State-1}, we consider the presence of heavy rain, which strongly affects the RF transmission and has fewer effects on FSO transmission. Thus, the RF link is not suitable for transmission, and the communication is conducted through the FSO link using the total transmit power is $P_1=P_t$. Likewise, in State-2, we consider that FSO is totally unavailable due to heavy fog and the RF transmission is activated using the total transmit power $P_2=P_t$.
Furthermore, we denote the probability of occurrence of each state by $p_k$ for $k\in \lbrace{0,1,2}\rbrace $.

\vspace{-0.5cm}
\subsection{Outage Probability}
To study the performance of our proposed scheme, we derive the OP expressions. The OP can be defined as the probability that the instantaneous SNR $\gamma$ falls below a predefined threshold $\gamma_{th}$ and it can be written for the $k$-th state as $P_{\text{out},k}(\gamma_{th})=\Pr[\gamma \leq \gamma_{th}]$ for $k \in \left\lbrace 0,1,2\right\rbrace$. For State-0, where both RF and FSO links might be active simultaneously, the GS uses SC. Thus, the final expression of OP for State-0 in the absence of pointing errors can be obtained as\footnote{The infinite summation in (\ref{Pout1}) converges very fast after 5 to 10 iterations with $10^{-6}$ convergence error.}

\vspace{-0.4cm}
\small
\begin{align}
\label{Pout1}
    &P_{\text{out},0} (\gamma_{th})= \sum_{\rho=0}^{\infty} \left( \begin{array}{c} \alpha \\
 \rho
   \end{array}  \right) 
   (-1)^{\rho} \exp\left[ -\rho \left( \frac{\gamma_{th}}{{\eta}^2 \frac{ P_1 (I^a)^2 }{N_{0}}}\right) ^{\frac{\beta}{2}}\right] \nonumber\\
   &\times \Bigg ( 1- \sum_{p=0}^{m -1} \sum_{q=0}^{p} \frac{\mu  (1-m)_p  (-\delta )^p }{q! \phi ^{p-q+1}(\frac{P_2 h_l}{N_0}) ^{p+1} p!} \gamma_{th}^q\exp(-\phi  \gamma_{th}) \Bigg),
\end{align}
\normalsize
where $(\cdot)_p$ is the Pochhammer symbol, $\delta =\frac{\Omega }{2b (2b  m  + \Omega )}$, and $\phi =\frac{\nu - \delta }{\overline{\gamma}_{RF}}$ \cite{2019physical}. {In the presence of non-zero boresight pointing errors, the OP for State-0 can be easily derived.}
On the contrary, the OP expressions for State-1 and State-2 can be simply obtained from their CDFs by using $P_{out}(\gamma_{th})=~F(\gamma_{th})$. In the presence of non-zero boresight pointing error on State-1, the OP can be expressed as 

\vspace{-0.17cm}
\small
\begin{align}
& P_{\text{out},1} (\gamma_{th})=\frac{\alpha g^2}{\beta} \exp(- \frac{s^2}{2\sigma_s^2}) \sum_{i=0}^{\infty} 
\frac{(-1)^i \Gamma(\alpha)}{(1+i)!\Gamma(\alpha-i)} \nonumber\\ 
&\times \sum_{j=0}^{\infty} \frac{(\frac{s^2 g^2}{2\sigma_s^2 \beta})^j}{j!}
G_{j+2,j+3}^{j+2,1} \Bigg( T_3'(i) \Big| \begin{array}{c}
1, \overbrace{T_1 +1,...,T_1+1}^{(j+1) \text{terms}} \\
1,\underbrace{T_1,...,T_1}_{(j+1) \text{terms}},0
\end{array} 
\Bigg),
\end{align}
\normalsize
where $T_3'(i)=(1+i) \Big( \frac{1}{\eta A_0} \sqrt{\frac{\gamma_{th}}{ P_t (I^a)^2/N_{0}}}\Big )^\beta$\footnote{The double infinite summations, require 3 to 5 terms to converge with a convergence error $ 2 \times10^{-7}$.}. In the absence of pointing error, the OP expression for State-1 can be given similarly to \cite[eqn. (23)]{erdogan2020}.
For State-2, the OP is given as

\vspace{-0.2cm}
\small
\begin{align}
P_{\text{out},2}(\gamma_{th})= 1- \sum_{p=0}^{m -1} \sum_{q=0}^{p} \frac{\mu (1-m)_p(-\delta )^p }{q! \phi ^{p-q+1}(\frac{P_t h_l}{N_0})^{p+1} p!} \gamma_{th}^q\exp(-\phi  \gamma_{th}).
\end{align}
\normalsize
Finally, the average performance of our proposed model can be given as $P_{out}(\gamma_{th})= \sum_{k=0}^{2} p_k P_{out,k}(\gamma_{th})$. 
%*********************************%%%
\subsection{High SNR Analysis
In this subsection, we focus on the high SNR analysis and obtain the diversity order to gain more insights about the system behavior. The diversity order defines the slope of the OP versus the average SNR at asymptotically high SNR \cite{ashrafzadeh2020unified}. 
To do so, for FSO transmission in the absence of pointing errors, we apply $\exp(-x/a) \simeq 1-x/a$ into the CDF of EW fading given in \cite{erdogan2020}. Thus, after few manipulations, the asymptotic OP can be obtained as 
\begin{align}
   P_{\text{out,1}}^{\infty}&= \left[ \left( \frac{\gamma_{th}}{{(\eta} I^a)^2 }\right) ^{\frac{\alpha\beta}{2}}\right] \left(\frac{1}{P_1/N_{0}}\right)^{\frac{\alpha\beta}{2}} =G_c (\overline{\gamma})^{-G_d^{FSO}},
\end{align}
where $G_c= \left( \frac{\gamma_{th}}{{(\eta}I^a)^2 }\right) ^{\frac{\alpha\beta}{2}}$ defines the shift of the curve in SNR and $ G_d^{FSO}=\frac{\alpha\beta }{2}$ presents the diversity order.}

For the RF transmission, we need to apply Maclaurin series expansion \cite{2014table} for the exponential function and consider only the first term as the higher-order terms are neglected. Accordingly, the OP at high SNR can be obtained as 
\begin{align}
 &P_{\text{out,2}}^{\infty} \simeq \mu \gamma_{th}\frac{ 1}{{P_2 h_l}/{N_0}}.
\end{align}
Thus, the diversity order is $G_d^{RF}=1$ as we use a single antenna. Therefore, for State-0, where SC is adopted on the GS, the diversity gain is $G_d=\max(\frac{\alpha\beta }{2}, 1)$.

%****************************%%
\section{Numerical Results and Discussions}
%****************************%%
\begin{figure}[!t]
  \centering
    \includegraphics[width=3.4in]{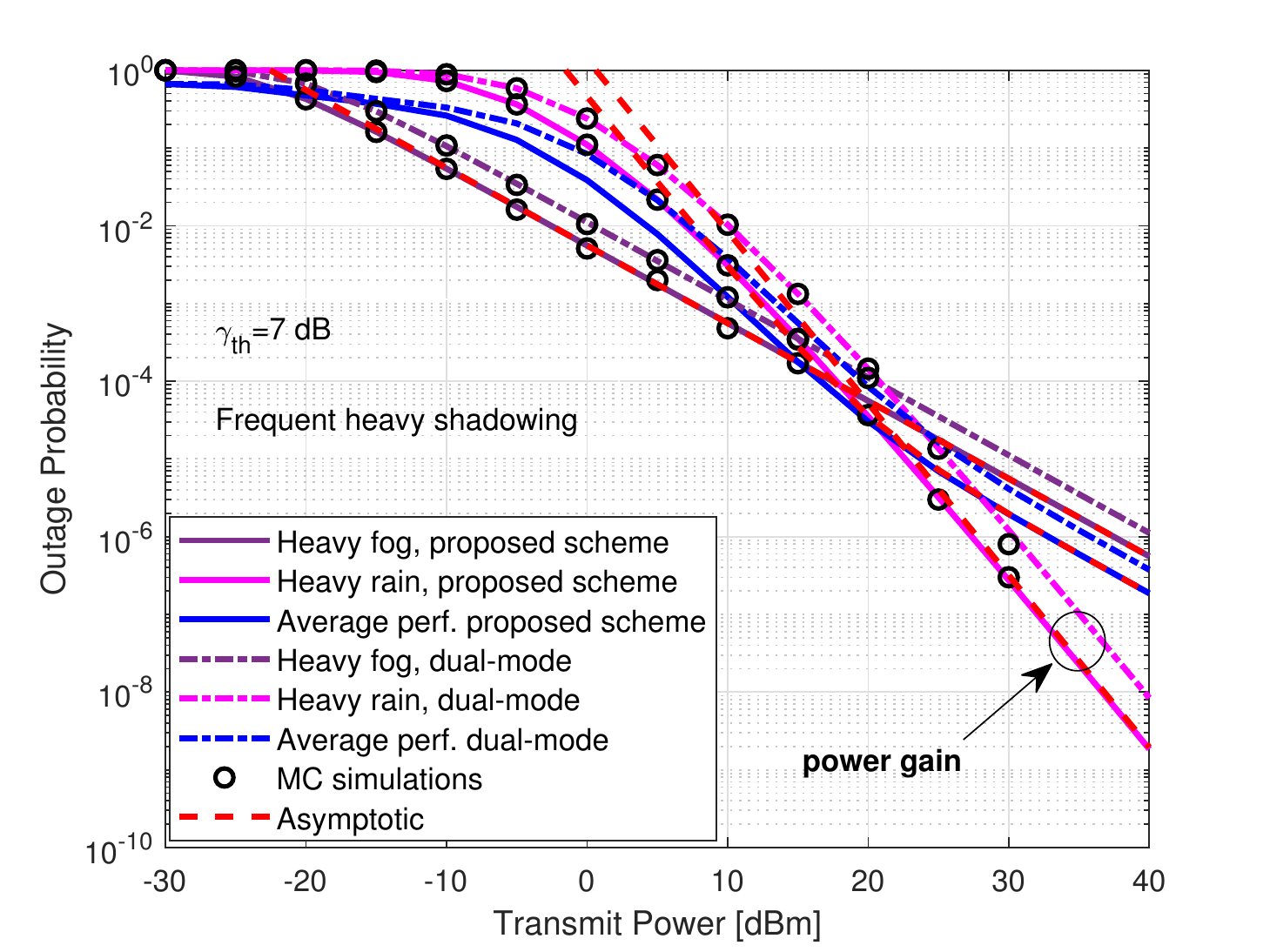}
    \caption{Outage probability performance of the proposed scheme.}
  \label{fig:OP}
\end{figure}
\begin{figure}[!t]
  \centering
    \includegraphics[width=3.4in]{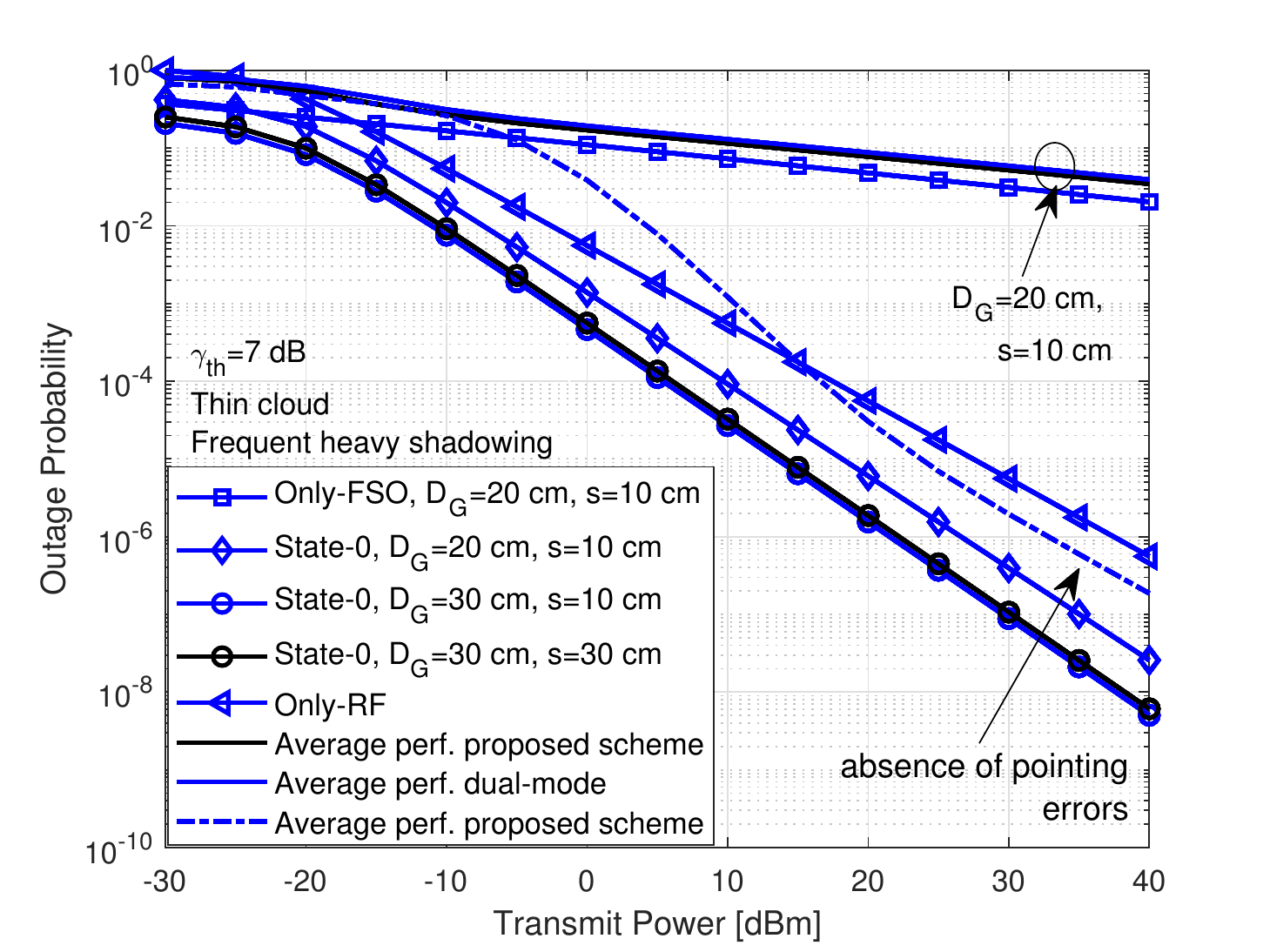}
    \caption{Outage probability performance of the proposed scheme in the presence of non-zero boresight pointing errors.}
  \label{fig:heavy rain}
\end{figure}
In this section, we evaluate the performance of our system in terms of outage probability. The satellite is assumed to be orbiting at an altitude of $500$ km, communicating with a GS located at $0.8$ km above sea level. For the FSO link, $\lambda =1550$ nm wavelength is considered, whereas the RF link is operating at $f=40$ GHz. For FSO communication, the zenith angle is set to $\xi=65$°, the nominal value of the refractive index is $C_0=5\times~10^{-14}$ \cite{kazemi}, and the wind speed is set to $v_g=2.8$ m/s for thin cirrus clouds. In the presence of heavy rain, we consider $C_0=0.4\times~10^{-14}$ \cite{kazemi}, $v_g=11.176$ m/s, and the specific attenuation of foggy weather is considered as $339.62$ dB/km. For the RF link, we consider the presence of frequent heavy shadowing where $m=1.0, b=0.063, \Omega=8.94 \times 10^{-4}$. Moreover, the transmitter and receiver gains are set to $G_T=G_R=45$ dB, the oxygen scattering parameter $\omega_{oxy}=0.1$ dB/km, and the rain rate is given as $\mathcal{R}=25$ mm/h. Furthermore, in the presence of non-zero boresight pointing errors, we set $\theta=0.2$ mrad, $\sigma_s=20$ cm and we consider different sizes of boresight and aperture. In all simulations, the threshold is set to $\gamma_{th}=7$ dB. Without loss of generality, we simply assume that the different states of weather have the same probability of occurrence satisfying $\sum_{k=0}^{2}p_{k}=1$. Thereafter, the probabilities are given as $p_k=1/3$ for $k={0,1,2}$.
In addition, we compare our proposed scheme with the conventional dual-mode hybrid RF/FSO, in which the total power is always divided into $P_t/2$ for all states. Finally, the derived expressions are verified using Monte Carlo (MC) simulations, however, not explicitly included to guarantee the clarity of the figure.

Fig. 2 illustrates the outage probability vs. transmit power for the proposed strategy in the absence of pointing errors. As we can see, the analytical results match the MC simulations with good accuracy. In addition, the high SNR analysis shows that for the heavy fog state, the overall gain is 1, whereas, for the heavy rain state, the overall gain is close to $\frac{\alpha \beta}{2}=2.2117$.
Moreover, as can be seen in the figure, at lower transmit power, the performance of foggy weather is better than the performance of rainy weather. However, after $P_t=16$ dBm, the performance of rainy weather improves, causing a crossing point between the two curves. We also compare our proposed strategy with the dual-mode scheme. Furthermore, the figure shows that the proposed strategy achieves better outage performance than the conventional dual-mode hybrid RF/FSO communication.

Fig. 3 investigates the impact of non-zero boresight pointing errors and the impact of aperture averaging for the proposed scheme. It is clear from the figure, that the average performance of the proposed scheme highly deteriorates in the presence of boresight non-zero pointing errors. As observed in the figure, for thin cloud conditions, the proposed hybrid RF/FSO configuration performs better than FSO-only and RF-only transmissions. In addition, in the presence of non-zero boresight pointing errors, we can observe that the proposed strategy performs better than the dual-mode hybrid communication. Moreover, we can see that RF-only scheme outperforms FSO-only scheme due to the huge impact of non-zero boresight pointing errors. Furthermore, the results demonstrate that increasing the aperture size $D_G$ enhances the overall performance by reducing the impact of atmospheric turbulence and pointing errors. Also, when the boresight displacement $s$ increases, the misalignment between the satellite and the GS increases, therefore, less power is collected by the receiver aperture yielding a serious degradation in the OP performance.

%the heavy rain state. As observed from the figure, pointing errors can cause misalignment, which degrades the overall outage performance of the proposed scheme. Also, as we can see, when the boresight displacement $s$ increases, the misalignment between the satellite and the GS increases, therefore, less power is collected by the receiver aperture yielding a serious depgradation in the OP performance. Furthermore, the results demonstrate that increasing the aperture size $D_G$ enhances the overall performance by reducing the impact of atmospheric turbulence. Finally, we can observe that the proposed strategy outperforms the dual-mode hybrid communication.
\balance
%****************************%%

In summary, we provide some important design guidelines that can be helpful for the design downlink hybrid RF/FSO SatCom systems.
\begin{itemize}
  \item We observe from the results that the proposed model is superior to the dual-mode hybrid RF/FSO communication method for SatCom. 
  \item In the absence of pointing errors, only at lower transmission power, the RF communication outperforms FSO communication. However, in the presence of non-zero boresight pointing errors, the RF transmission performs better than FSO transmission. 
\item The results have shown that the non-zero boresight pointing error strongly deteriorates the overall performance of FSO communication as it increases the misalignment between communicating parts. 
\item It is shown in Fig. 3 that the aperture averaging technique is an effective solution in SatCom to mitigate the effects of atmospheric turbulence and pointing errors. 
\end{itemize}

%****************************%%
\section{Conclusion}
%****************************%%
In this paper, we have proposed a new soft-switching setup for hybrid RF/FSO downlink SatCom. According to the context-aware sensor system, an LEO satellite adapts its power and switches between RF and FSO links by selecting the channel with the highest performance based on the weather conditions. To quantify the performance of our model, outage probability expressions are derived and simulation results are provided to confirm the accuracy of our analytical results. Moreover, the results have shown that our proposed strategy improves power efficiency compared to the conventional dual-mode hybrid RF/FSO. Furthermore, we have studied the impact of non-zero boresight pointing errors to observe the misalignment effect for hybrid RF/FSO communication. The study indicates that increasing the boresight leads to severe performance degradation. However, the aperture averaging technique can be used to improve the performance and alleviate pointing error effects. Finally, some significant design guidelines are suggested.

%\addtolength{\topmargin}{0in}
%\vspace{-0.3cm}
 \balance
\bibliographystyle{IEEEtran}
{\linespread{0.95}\bibliography{ref}}

% Generated by IEEEtran.bst, version: 1.14 (2015/08/26)
\begin{thebibliography}{10}
\providecommand{\url}[1]{#1}
\csname url@samestyle\endcsname
\providecommand{\newblock}{\relax}
\providecommand{\bibinfo}[2]{#2}
\providecommand{\BIBentrySTDinterwordspacing}{\spaceskip=0pt\relax}
\providecommand{\BIBentryALTinterwordstretchfactor}{4}
\providecommand{\BIBentryALTinterwordspacing}{\spaceskip=\fontdimen2\font plus
\BIBentryALTinterwordstretchfactor\fontdimen3\font minus
  \fontdimen4\font\relax}
\providecommand{\BIBforeignlanguage}[2]{{%
\expandafter\ifx\csname l@#1\endcsname\relax
\typeout{** WARNING: IEEEtran.bst: No hyphenation pattern has been}%
\typeout{** loaded for the language `#1'. Using the pattern for}%
\typeout{** the default language instead.}%
\else
\language=\csname l@#1\endcsname
\fi
#2}}
\providecommand{\BIBdecl}{\relax}
\BIBdecl

\bibitem{9210567}
O.~Kodheli, E.~Lagunas, N.~Maturo, S.~K. Sharma, B.~Shankar, J.~F.~M. Montoya,
  J.~C.~M. Duncan, D.~Spano, S.~Chatzinotas, S.~Kisseleff, J.~Querol, L.~Lei,
  T.~X. Vu, and G.~Goussetis, ``Satellite communications in the new space era:
  A survey and future challenges,'' \emph{IEEE Communications Surveys
  Tutorials}, vol.~23, no.~1, pp. 70--109, 2021.

\bibitem{9296829}
E.~Erdogan, I.~Altunbas, N.~Kabaoglu, and H.~Yanikomeroglu, ``A cognitive radio
  enabled {RF/FSO} communication model for aerial relay networks: Possible
  configurations and opportunities,'' \emph{IEEE Open Journal of Vehicular
  Technology}, vol.~2, pp. 45--53, 2021.

\bibitem{viswanath2015}
A.~Viswanath, V.~K. Jain, and S.~Kar, ``Analysis of {E}arth-to-satellite
  free-space optical link performance in the presence of turbulence,
  beam-wander induced pointing error and weather conditions for different
  intensity modulation schemes,'' \emph{IET Comm.}, vol.~9, no.~18, pp.
  2253--2258, 2015.

\bibitem{2018outage}
E.~T. Michailidis, N.~Nomikos, P.~Bithas, D.~Vouyioukas, and A.~G. Kanatas,
  ``Outage probability of triple-hop mixed {RF/FSO/RF} stratospheric
  communication systems,'' in \emph{IEEE Int. Conf. on Adv. in Satellite and
  Space Commun.(SPACOMM)}, 2018, pp. 1--6.

\bibitem{touati}
A.~Touati, A.~Abdaoui, F.~Touati, M.~Uysal, and A.~Bouallegue, ``On the effects
  of combined atmospheric fading and misalignment on the hybrid {FSO/RF}
  transmission,'' \emph{J. of Optical Commun. and Netw.}, vol.~8, no.~10, pp.
  715--725, 2016.

\bibitem{soleimani2015generalized}
E.~Soleimani-Nasab and M.~Uysal, ``Generalized performance analysis of mixed
  {RF/FSO} cooperative systems,'' \emph{IEEE Transactions on Wireless
  Communications}, vol.~15, no.~1, pp. 714--727, 2015.

\bibitem{ashrafzadeh2020unified}
B.~Ashrafzadeh, A.~Zaimbashi, E.~Soleimani-Nasab, and M.~Uysal, ``Unified
  performance analysis of multi-hop {FSO} systems over double generalized gamma
  turbulence channels with pointing errors,'' \emph{IEEE Transactions on
  Wireless Communications}, vol.~19, no.~11, pp. 7732--7746, 2020.

\bibitem{weather}
F.~Nadeem, V.~Kvicera, M.~S. Awan, E.~Leitgeb, S.~S. Muhammad, and G.~Kandus,
  ``Weather effects on hybrid {FSO/RF} communication link,'' \emph{IEEE {J}. on
  {S}el. {A}reas in {C}ommun.}, vol.~27, no.~9, pp. 1687--1697, 2009.

\bibitem{kazemi}
H.~Kazemi, M.~Uysal, and F.~Touati, ``Outage analysis of hybrid {FSO/RF}
  systems based on finite-state {M}arkov chain modeling,'' \emph{Int. Workshop
  in Optical Wireless Commun. (IWOW)}, pp. 11--15, 2014.

\bibitem{siddharth2020}
M.~Siddharth, S.~Shah, and S.~R., ``Outage analysis of adaptive combining
  scheme for hybrid {FSO/RF} communication,'' in \emph{IEEE National Conf. on
  Commun. (NCC)}, 2020, pp. 1--6.

\bibitem{9446153}
S.~Shah, M.~Siddharth, N.~Vishwakarma, R.~Swaminathan, and A.~S. Madhukumar,
  ``Adaptive-combining-based hybrid {FSO/RF} satellite communication with and
  without {HAPS},'' \emph{IEEE Access}, vol.~9, pp. 81\,492--81\,511, 2021.

\bibitem{swaminathan2021haps}
S.~R, S.~Sharma, N.~Vishwakarma, and A.~S. Madhukumar, ``{HAPS}-based relaying
  for integrated space-air-ground networks with hybrid {FSO/RF} communication:
  A performance analysis,'' \emph{IEEE Transactions on Aerospace and Electronic
  Systems}, vol.~57, no.~3, pp. 1581--1599, 2021.

\bibitem{ben2021haps}
O.~B. Yahia, E.~Erdogan, G.~K. Kurt, I.~Altunbas, and H.~Yanikomeroglu,
  ``{HAPS} selection for hybrid {RF/FSO} satellite networks,'' \emph{arXiv
  preprint arXiv:2107.12638}, 2021.

\bibitem{barrios2013}
R.~Barrios~Porras, \emph{Exponentiated Weibull Fading Channel Model in
  Free-Space Optical Communications under Atmospheric Turbulence}.\hskip 1em
  plus 0.5em minus 0.4em\relax Ph.D. dissertation, Dept. Signal Theory Commun.,
  Univ. Polit{\`e}cnica de Catalunya ({UPC}), {B}arcelona, {S}pain, May 2013.

\bibitem{yi2015free}
X.~Yi and M.~Yao, ``Free-space communications over exponentiated weibull
  turbulence channels with nonzero boresight pointing errors,'' \emph{Optics
  Express}, vol.~23, no.~3, pp. 2904--2917, 2015.

\bibitem{erdogan2020}
E.~{Erdogan}, I.~{Altunbas}, G.~K. {Kurt}, M.~{Bellemare}, G.~{Lamontagne}, and
  H.~{Yanikomeroglu}, ``Site diversity in downlink optical satellite networks
  through ground station selection,'' \emph{IEEE Access}, vol.~9, pp.
  31\,179--31\,190, 2021.

\bibitem{2018performance}
Y.~Wang, P.~Wang, X.~Liu, and T.~Cao, ``On the performance of dual-hop mixed
  {RF/FSO} wireless communication system in urban area over aggregated
  exponentiated {W}eibull fading channels with pointing errors,'' \emph{Optics
  Commun.}, vol. 410, pp. 609--616, 2018.

\bibitem{yang2014free}
F.~Yang, J.~Cheng, and T.~A. Tsiftsis, ``Free-space optical communication with
  nonzero boresight pointing errors,'' \emph{IEEE Trans. on Commun.}, vol.~62,
  no.~2, pp. 713--725, 2014.

\bibitem{Wolform}
\BIBentryALTinterwordspacing
The {W}olfram functions site. [Online]. Available: \url{http://www.wolfram.com}
\BIBentrySTDinterwordspacing

\bibitem{6605500}
M.~R. Bhatnagar and M.~K. Arti, ``Performance analysis of hybrid
  satellite-terrestrial {FSO} cooperative system,'' \emph{IEEE Photonics
  Technology Letters}, vol.~25, no.~22, pp. 2197--2200, 2013.

\bibitem{ITUrain}
ITU, ``Specific attenuation model for rain for use in prediction
  methods.''\hskip 1em plus 0.5em minus 0.4em\relax Recommendation P.838-3,
  2003.

\bibitem{paris2010closed}
J.~F. Paris, ``Closed-form expressions for {R}ician shadowed cumulative
  distribution function,'' \emph{Electronics Letters}, vol.~46, no.~13, pp.
  952--953, 2010.

\bibitem{2014table}
I.~S. Gradshteyn and I.~M. Ryzhik, \emph{Table of Integrals, Series, and
  Products}.\hskip 1em plus 0.5em minus 0.4em\relax Academic Press, 2014.

\bibitem{2019physical}
Y.~Ai, A.~Mathur, M.~Cheffena, M.~R. Bhatnagar, and H.~Lei, ``Physical layer
  security of hybrid satellite-{FSO} cooperative systems,'' \emph{IEEE Photon.
  J.}, vol.~11, no.~1, pp. 1--14, 2019.

\end{thebibliography}
%\vspace{-0.1cm}
%\bibliography{ref}
\end{document}